\documentclass{ws-procs9x6-cpt22}
\begin{document}

\newcommand{\refeq}[1]{(\ref{#1})}

\title{Lorentz Violating Electrodynamics with sources and mirrors}

\author{A. F. Ferrari$^1$}

\address{$^1$Universidade Federal do ABC - UFABC, Centro de Ciências Naturais e
Humanas, Rua Santa Adélia, 166, 09210-170, Santo André, SP, Brazi}

\begin{abstract}
We look for modifications that can be derived from the Maxwell sector
of the SME for the interaction between electrical charges and the
electromagnetic field. The kind of problem that we are looking into
is the one that might be found in a standard textbook on electromagnetism:
given a set of electrical charges, in the presence of some non-trivial
boundary condition (such as a mirror), find the electromagnetic field
and the interaction energy between those charges. To study such kind
of problem in the presence of LV coefficients, we calculate the modified
propagator for the Maxwell field, and use it to calculate
energy of a system given by an semi-transparent mirror and a point-like
charge. We show how new effects appear in this case, such as a spontaneous
torque acting on the system.
\end{abstract}

\bodymatter

\phantom{}\vskip10pt\noindent
One of the most studied sectors of the
SME\,\citep{SME1,SME2} is certainly the Maxwell sector, one of the
reasons being the huge importance of experiments involving photons
in establishing constraints on LV\,\citep{VANR}. We refer the reader
to our paper\,\citep{Borges2022} for an extensive, yet not exhaustive
list of references regarding both classical and quantum electrodynamics with LV.
Our interest is to extend these kind of studies including the presence
of nontrivial boundary conditions. There are many applications of
nontrivial boundary conditions in field theory, such as the study
of the Casimir effect, and again we refer to\,\citep{Borges2022}
for a list of references for the interested reader. 

The specific problem
we describe in this work if a study of classical electrodynamics with
Lorentz symmetry breaking in the presence of a semi-transparent mirror.
This is an interesting topic since, in practice, electromagnetic configurations
in actual experiments do usually involve conductors, which have to
be properly considered in the theoretical models. 

We choose to restrict ourselves to the context of the nonbirefringent
CPT-even pure-photon sector of the SME. 
We first consider, for simplicity, the case when the LV background
is described by a single vector $v^{\mu}$, but we are able to generalize
our results afterwards. Throughout the work we work in a $3+1$-dimensional
Minkowski space-time with metric $\eta^{\rho\nu}=(1,-1,-1,-1)$. The
Levi-Civita tensor is denoted by $\epsilon^{\rho\nu\alpha\beta}$
with $\epsilon^{0123}=1$.

The presence of a semi-transparent boundary or a two-dimensional semi-transparent
mirror (which, with no loss of generality, can be considered as perpendicular
to the $x^{3}$ axis, located at $x^{3}=a$) can be represented by
a $\delta$-like term, that is to say,
\begin{eqnarray}
{\cal {L}}&=&-\frac{1}{4}F_{\mu\nu}F^{\mu\nu}-\frac{1}{2\gamma}\left(\partial_{\mu}A^{\mu}\right)^{2}-\frac{1}{2}v^{\mu}v_{\nu}F_{\mu\lambda}F^{\nu\lambda}\nonumber\\
&&-\frac{1}{m}\left(\frac{1}{2}S^{\mu}\epsilon_{\mu\nu\alpha\beta}F^{\alpha\beta}\right)^{2}\delta\left(x^{3}-a\right)-J^{\mu}A_{\mu}\ .\label{LVMirror}
\end{eqnarray}
where $A^{\mu}$ is the gauge field, $F^{\mu\nu}=\partial^{\mu}A^{\nu}-\partial^{\nu}A^{\mu}$
is the field strength, $\gamma$ is a gauge fixing parameter, $j^{\mu}$
is an external source, $S^{\gamma}=\eta_{\ 3}^{\gamma}$ is the vector
normal to the mirror and $m^{-1}>0$ is a coupling constant with inverse
of mass dimension, establishing the degree of transparency of the
mirror, the limit $m\rightarrow0$ corresponding to a perfect mirror\,\citep{FABFEB}.
The background vector $v^{\mu}$ appearing here is contained in the
$\left(K_{F}\right)^{\mu\nu\alpha\beta}$ as described in\,\citep{SME2}.
This very particular choice allows us to perform all calculations
analytically, in perturbation theory, up to the second order of $v^{\mu}$.

The strategy is quite simple: from\,\eqref{LVMirror} we obtain a
modified propagator, which is then used to calculate the interaction
energy between electromagnetic sources via 
\begin{equation}
E=\frac{1}{2T}\int d^{4}x\ d^{4}y\ J^{\mu}\left(x\right){G}_{\mu\nu}\left(x,y\right)J^{\nu}\left(y\right)\ ,\label{energy}
\end{equation}
where $T$ is a time interval, and it is implicit the limit $T\to\infty$
at the end of the calculations. 
The calculation itself is quite involved
as can be seen in\,\citep{Borges2022}. The modified propagator,
for example, is given as 
\begin{equation}
G_{\mu\nu}\left(x,y\right)=G_{\mu\nu}^{(0)}\left(x,y\right)+{\bar{G}}_{\mu\nu}\left(x,y\right)\ ,\label{prop9}
\end{equation}
where $G_{\mu\nu}^{(0)}$ is the unmodified Lorentz invariant part,
\begin{eqnarray}
G_{\mu\nu}^{(0)}\left(x,y\right)&=&-\int\frac{d^{4}p}{(2\pi)^{4}}\frac{e^{-ip\cdot(x-y)}}{p^{2}} \times \nonumber\\
&&\times \Biggl[\left(1-\frac{(p\cdot v)^{2}}{p^{2}}\right)\eta_{\mu\nu}-v_{\mu}v_{\nu}+\frac{(p\cdot v)}{p^{2}}(p_{\mu}v_{\nu}+v_{\mu}p_{\nu})\Biggr]\ ,\label{prop0}
\end{eqnarray}
while the modified part ${\bar{G}}$ is a cumbersome expression that can be seen in\,\citep{Borges2022}.

Despite the complexity of these expressions, we are able to calculate
the interaction energy between a point like charge located at the
position ${\bf b}=\left(0,0,b\right)$ and the mirror, with the result
\begin{eqnarray}
E_{MC}&=&-\frac{q^{2}}{16\pi R}\Biggl\{1-2mRe^{2mR}Ei\left(1,2mR\right)\nonumber\\
&&+\left[{\bf {v}}_{\parallel}^{2}-2\left(v^{0}\right)^{2}\right]\left[\frac{1}{2}-mR+2\left(mR\right)^{2}e^{2mR}Ei\left(1,2mR\right)\right]\Biggr\}\, ,\label{energyC3}
\end{eqnarray}
with $v_{\parallel}^{\mu}=\left(v^{0},v^{1},v^{2}\right)$, and $v^{3}$
standing for the background vector parallel and perpendicular to the
mirror, respectively. Here, $Ei(n,s)$ is the exponential integral function\,\citep{Arfken}.

When we fix the distance between the charge and the mirror, from Eq.
(\ref{energyC3}), we see that the whole system feels a torque depending
on its orientation with respect to the background vector. In order
to calculate this torque, we define as $0\leq\alpha\leq\pi$ the angle
between the normal to the mirror and the background vector ${\bf {v}}$,
in such a way that, ${\bf {v}}_{\parallel}^{2}={\bf {v}}^{2}\sin^{2}\left(\alpha\right)$.
This leads to
\begin{equation}
\tau_{MC}=-\frac{\partial E_{MC}}{\partial\alpha}=\frac{q^{2}{\bf {v}}^{2}}{16\pi R}\sin\left(2\alpha\right)\left[\frac{1}{2}-mR+2\left(mR\right)^{2}e^{2mR}Ei\left(1,2mR\right)\right]\ .\label{Torque}
\end{equation}

These results can be extended for more general LV settings by using
the image method and matching to the specific case we calculated
explicitly. What we obtain is
\begin{align}
E'_{MC} & =-\frac{q^{2}}{16\pi R}\Biggl[1-2mRe^{2mR}Ei\left(1,2mR\right)\label{EMCKF12}\\
 & +2\left[\left(K_{F}\right)^{0101}+\left(K_{F}\right)^{0202}\right]{\cal {F}}_{2}\left(mR\right)\Biggr]\ .
\end{align}
Notice this is still for a mirror perpendicular to the $x^{3}$. From
here one can also derive a spontaneous torque acting on the system. 

We can estimate the order of magnitude of this generated torque in
the following way: we consider a typical distance of atomic experiments
in the vicinity of conductors (mirrors) of order $R\sim10^{-6}$m,
the fundamental electronic charge $q\sim1.60217\times10^{-19}\text{C}$,
as well as the upper bounds ${\tilde{\kappa}}_{\mathrm{tr}}\sim1.4\times10^{-19}$,
$\left({\tilde{\kappa}}_{e-}\right)^{ij}\sim4\times10^{-18}$ and
$\left({\tilde{\kappa}}_{e+}\right)^{ij}\sim2\times10^{-37}$ obtained
from \citep{VANR,FRKlin}. In this case, for a perfect mirror, corresponding
to the limit $m\rightarrow0$, we have $\tau{}_{MC}\sim10^{-40}$Nm.
For an imperfect mirror, the magnitude of the torque is smaller. Taking
$m\sim10^{-5}\text{GeV}$, for example, we obtain $\tau_{MC}\sim10^{-45}\,\text{Nm}$.
This effect is out of reach of being measured by using current technology.

In summary, we were able to exhibit new effects that would arise if we put a single point like
charge in the presence of an imperfect mirror, due to the presence of nonbirrefringent LV
described by a $K_F$ term. These effects, however, are two small to expect them to provide any interesting
new constraint. 

\section*{Acknowledgments}

This study was financed in part by the Coordenação de Aperfeiçoamento
de Pessoal de Nível Superior -- Brasil (CAPES) -- Finance Code 001
(LHCB), and Conselho Nacional de Desenvolvimento Científico e Tecnológico
(CNPq) via the grant 305967/2020-7 (AFF).


\begin{thebibliography}{99}
\bibitem{SME1} D. Colladay, V. A. Kostelecký, Phys. Rev. D \textbf{{55}}, 6760 (1997).

\bibitem{SME2} D. Colladay, V. A. Kostelecký, Phys. Rev. D \textbf{{58}}, 116002 (1998).

\bibitem{VANR} V.A. Kostelecký, N. Russell, Rev. Mod. Phys. \textbf{83}, 11 (2011) doi:10.1103/RevModPhys.83.11 {[}arXiv:0801.0287v15 (2022){]}.

\bibitem{Borges2022}L. H. C. Borges, A. F. Ferrari, Nucl. Phys. B \textbf{980}, 115829 (2022) doi:10.1016/j.nuclphysb.2022.115829 {[}arXiv:2205.05542{]}.

\bibitem{FABFEB} F.A. Barone and F.E. Barone, Phys. Rev. D \textbf{{89}}, 065020 (2014).

\bibitem{Arfken} G.B. Arfken, H.J. Weber, \textit{{Mathematical Methods for Physicists}} (Academic Press, USA, 1995).

\bibitem{FRKlin} F.R. Klinkhamer and M. Risse, Phys. Rev. D \textbf{{77}}, 117901 (2008).
\end{thebibliography}
\end{document}